MATTERS ARISING

# Is $p$-type doping in TeO$_2$ feasible?

Zewen Xiao,[1,*] Chen Qiu,[2] Su-Huai Wei,[2] and Hideo Hosono[3]

[1] Wuhan National Laboratory for Optoelectronic, Huazhong University of Science and Technology, Wuhan 430074, China

[2] Eastern Institute of Technology, Ningbo 315200, China

[3] MDX Research Center for Element Strategy, International Research Frontiers Initiative, Tokyo Institute of Technology, Yokohama 226-8501, Japan

Email: zwxiao@hust.edu.cn

ARISING FROM A. Zavabeti et al. *Nature Electronics* https://doi.org/10.1038/s41928-021-00561-5 (2021)

High-performance transparent oxide semiconductors (TOSs) are crucial for the advancement of next-generation transparent electronics, power electronics and energy-efficient displays. The lack of high-mobility, *p*-type TOSs presents a significant challenge in designing bipolar transistors, inverter circuits, and transparent thin-film transistors (TFTs). This challenge arises from the deep and localized valence band maxima (VBMs) primarily composed of O 2p orbitals, hindering the generation and transport of holes.[1] To address this issue, a common strategy involves elevating and dispersing the VBM by utilizing cations with high-lying occupied $d^{10}$ or $s^2$ orbitals that can effectively hybridize with O 2p orbitals.[2] Notably, the use of Cu(I) with the highest-lying $d^{10}$ orbitals has facilitated the realization of Cu(I)-based *p*-type TOSs, exemplified by $CuAlO_2$.[2,3] On the other hand, cations with $s^2$ states conform to a specific atomic energy level sequence: Sn(II) → Pb(II) → Sb(III) → Bi(III) → Te(IV) → Po(IV).[4] For instance, SnO is recognized as a *p*-type semiconductor with a high-lying s-like VBM at –5.8 eV relative to the vacuum level.[5] In contrast, PbO encounters challenges in hole doping due to its deeper $6s^2$ states induced by relativistic effects,[4] resulting in a VBM at –6.2 eV.[6] Following this trend, it is anticipated that $Sb_2O_3$, $Bi_2O_3$, $TeO_2$, and $PoO_2$ may encounter similar obstacles in serving as *p*-type TOSs.

Interestingly, in their Article, Zavapeti et al.[7] presented a remarkable finding showcasing two-dimensional (2D) β-$TeO_2$ as a high-mobility *p*-type TOS with a 2D hole density ($p_{2D}$) of $1.04×10^9$ cm$^{-2}$ and an impressive hole mobility of 141 cm$^2$ V$^{-1}$ s$^{-1}$. This discovery challenges the conventional perception of $TeO_2$ as an insulator[8] and contradicts the anticipated difficulty in hole doping according to established trends. Notably, Zavabeti et al.[7] observed that the Fermi level ($E_F$) of 2D β-$TeO_2$ is situated approximately 0.9 eV above the VBM (refer to Fig. 3b in Ref. 7). Calculation using the formula $p_{2D} = N_{v,2D} \exp[-(E_F-E_v)/k_B T]$ (where $N_{v,2D}$ represents the effective density of states in the valence band, typically around $10^{12}$ cm$^{-2}$, $k_B$ is the Boltzmann constant, and $T$ is the absolute temperature) suggests that with the reported $E_F$, the $p_{2D}$ value at room temperature is negligible, less than 1 cm$^{-2}$ (essentially zero). This implies insulating properties for 2D β-$TeO_2$, which sharply contrasts with the documented *p*-type conductivity demonstrating a $p_{2D}$ value of $1.04×10^9$ cm$^{-2}$.

Zavabeti et al. synthesized the 2D β-$TeO_2$ by high-temperature surface oxidation of a eutectic mixture containing Te (5 wt%) and Se (95 wt%). Despite their efforts, residual Se was identified in the 2D β-TeO$_2$ samples due to Se's lower oxophilicity relative to Te (refer to Extended Fig. 3 in Ref. 7). Furthermore, the potential reduction of an extremely thin layer of β-TeO$_2$ to elemental Te remains a plausible event. TeO$_2$ is known to undergo partial decomposition either spontaneously or upon heating, resulting the generation of tellurium suboxide (TeO$_{2-x}$, where $x$ typically ranges from 0.7 to 0.9) and elemental Te.[9] Moreover, when subjected to an applied voltage of several volts, the intense electric field strength across the ultrathin layer can easily surpass the breakdown voltage, leading to the formation of elemental Te. Notably, elemental Se, elemental Te, and Te$_{1-x}$Se$_x$ alloy are all well-known high-mobility $p$-type semiconductors with three-dimensional (3D) hole densities ($p_{3D}$) on the order of 10$^{14}$ cm$^{-3}$ (equivalent to $p_{2D}$ on the order of 10$^9$ cm$^{-2}$) and high hole mobilities ranging from hundreds to over a thousand cm$^2$ V$^{-1}$ s$^{-1}$,[10–12] comparable to those observed in the 2D β-TeO$_2$ samples. Thus, it is crucial to elucidate the true source of the $p$-type conductivity in the 2D β-TeO$_2$ samples, whether originating from the 2D β-TeO$_2$ itself, residual elemental Se, reduced elemental or even Te$_{1-x}$Se$_x$, warranting further investigation.

The discussion now shifts towards assessing the dopability of TeO$_2$ to enhance our comprehension of its electrical properties. The dopability of a semiconductor can be effectively evaluated from its band edge positions.[13] Generally, a semiconductor is more readily doped with electrons when its conduction band minimum (CBM) is deeper, and with holes when its VBM is shallower. Semiconductors prone to electron doping (e.g., ZnO, In$_2$O$_3$, and SnO$_2$) typically have a CBM deeper than approximately –4.0 eV, while semiconductors amenable to hole doping (e.g., SnO, NiO, and CuAlO$_2$) generally exhibit a VBM shallower than about –6.0 eV.[2] These empirical benchmarks have steered the investigation of transparent semiconductors, particularly those of the $p$-type variety. For instance, contrary to expectations, PbO, a $p$-type SnO homolog, encountered impediments in hole doping due to its VBM (–6.2 eV), surpassing the empirical VBM threshold.[6] Concerning TeO$_2$, which manifests in three polymorphs (α-, β-, and γ-TeO$_2$) as illustrated in Supplementary Fig. 1, first-principles calculations (refer to computational methods in Supplementary Information) unveil that the CBMs are shallower than the empirical CBM threshold of –4.0 eV, while the VBMs are deeper than the empirical VBM threshold of –6.0 eV, as depicted in Fig. 1. This signals challenges in carrier doping for α-, β-, and γ-TeO$_2$, aligning with their recognized insulating traits. In contrast to 3D β-TeO$_2$, the 2D β-

TeO$_2$ monolayer displays a deeper VBM and a shallower CBM, posing increased hurdles in carrier doping. The deep VBM and the ensuing difficulty in achieving *p*-type doping in TeO$_2$ can be attributed to the relatively deep Te 5s$^2$ orbital (Supplementary Table 2), which exhibits insufficient hybridization with O 2p orbitals (Supplementary Fig. 2) and therefore cannot elevate the VBM as effectively as the high-lying Cu 3d$^{10}$ and Sn 5s$^2$ orbitals.

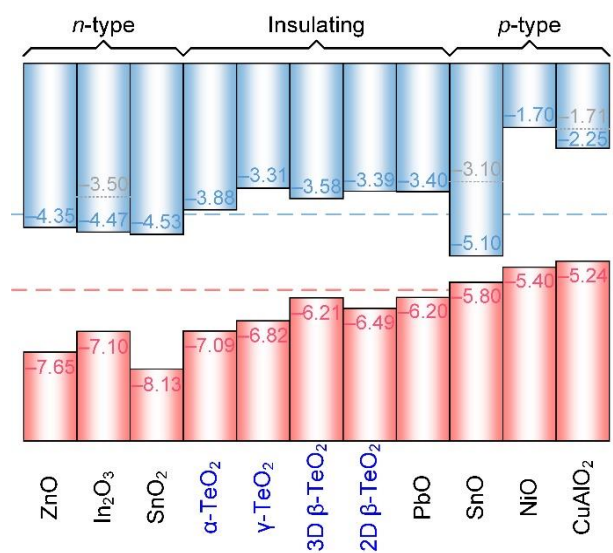

**Fig. 1 | Band alignment of TeO$_2$ polymorphs and related oxides.** The energies are with respect to the $E_{vac}$. The data for the TeO$_2$ polymorphs is theoretical, while the data for other compounds is experimental (Supplementary Table 1). Empirically, semiconductors with CBM deeper than the blue dashed line (–4 eV) can be doped *n*-type, while semiconductors with VBM shallower than the red dashed line (–6 eV) can be doped *p*-type.

The dopability of TeO$_2$ can also be elucidated by examining the thermodynamics of its intrinsic defects. Fig. 2 presents the calculated defect formation energies of both 3D and 2D β-TeO$_2$. Notably, all intrinsic defects exhibit relatively high formation energies, approaching or exceeding 1 eV at equilibrium, with deep transition levels (also depicted in Supplementary Fig. 3), signifying the absence of effective donors and acceptors. For 3D β-TeO$_2$ synthesized at room temperature, the calculated equilibrium $E_F$ ($E_{F,e}$) is situated at mid-gap (consistent across chemical conditions, as displayed in Fig. 2a and 2b), correlating with zero carrier density (refer to Supplementary Table 3). Elevating the growth temperature ($T_G$) to a sufficiently high temperature of 700 K marginally shift the $E_{F,e}$ away from mid-gap; however, the carrier densities (electron density $n_0$ of 8.2×10$^5$ cm$^{-3}$ under the Te-rich condition, $p_0$ = 3.1×10$^{10}$ cm$^{-3}$ under the O-rich condition) remain insufficient for typical semiconductor behavior. In the case of 2D β-TeO$_2$, the calculated $E_{F,e}$ consistently resides at mid-gap, irrespective of chemical conditions and $T_G$, as demonstrated in Fig. 2c and 2d. Moreover, donors and acceptors will spontaneously form in the *p*-type and *n*-type $E_F$ regions, respectively, owning to their negative formation energies, rendering external doping ineffective in altering its insulating traits.

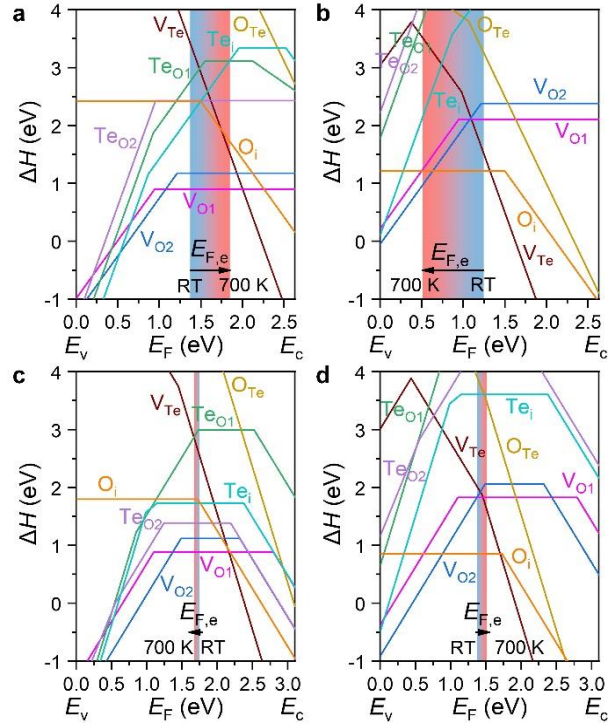

**Fig. 2 | Defect formation energies and equilibrium $E_F$.** Calculated formation energies (Δ$H$) of intrinsic defects as a function of the Fermi level ($E_F$) for 3D β-TeO$_2$ (**a,b**) and 2D β-TeO$_2$ (**c,d**) under Te-rich (**a,c**) and O-rich (**b,d**) conditions. The colored bars represent the ranges of equilibrium $E_F$ ($E_{F,e}$) solved with growth temperatures ranging from room temperature to a sufficient high temperature of 700 K.

In summary, our investigation unveils that TeO$_2$, whether in 3D bulk forms (α-, β-, or γ-TeO$_2$) or in 2D β-TeO$_2$ layers, encounters substantial obstacles in electrons and hole doping due to its excessively shallow CBM and excessively deep VBM, respectively, causing it to inherently exhibit insulating characteristics. Consequently, the observed *p*-type conductivity in the 2D β-TeO$_2$ samples by Zavabeti et al. likely stems from residual elemental Se, reduced elemental Te, or even Te$_{1-x}$Se$_x$ alloy, all of which function as high-mobility *p*-type semiconductors.

**Acknowledgements**

This work was supported by the National Natural Science Foundation of China (grants no. 52372150, 12088101, and 11991060) and the National Key R&D Program of China (2022YFB4200305).

**Author contributions**

Z.X. conceptualized the idea and performed the calculations. Z.X., C.Q., S.-H.W., and H.H. interpreted the results and wrote the paper.

**Competing interests**

The authors declares no competing interests.

Supplementary Information for:

MATTERS ARISING

# Is *p*-type doping in TeO$_2$ feasible?

Zewen Xiao,[1,*] Chen Qiu,[2] Su-Huai Wei,[2] and Hideo Hosono[3]

[1]Wuhan National Laboratory for Optoelectronic, Huazhong University of Science and Technology, Wuhan 430074, China

[2]Eastern Institute of Technology, Ningbo 315200, China

[3]MDX Research Center for Element Strategy, International Research Frontiers Initiative, Tokyo Institute of Technology, Yokohama 226-8501, Japan

Email: zwxiao@hust.edu.cn

ARISING FROM A. Zavabeti et al. *Nature Electronics* https://doi.org/10.1038/s41928-021-00561-5 (2021)

Contents:

Supplementary Fig. 1–3
Supplementary Table 1-3
Computational methods
Supplementary references

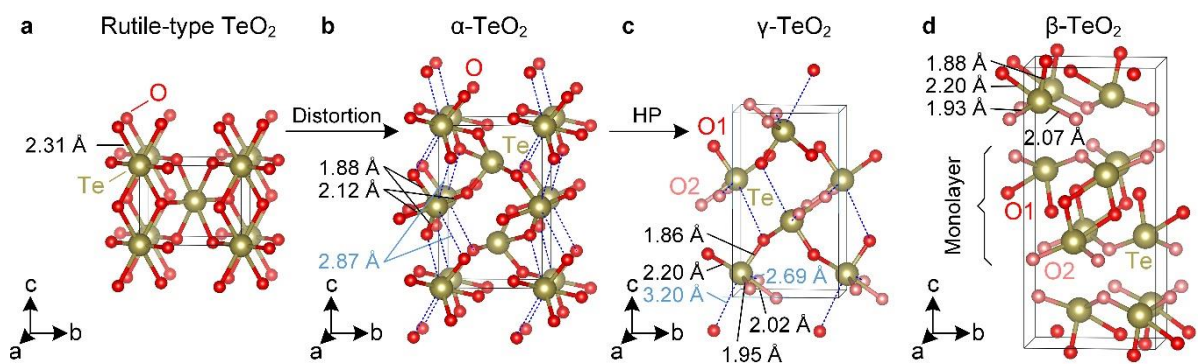

**Supplementary Fig. 1 | Crystal of TeO$_2$ polymorphs: a**, hypothetical rutile-type TeO$_2$ (space group $P4_2/mnm$), **b**, paratellurite α-TeO$_2$ (space group $P4_12_12$), **c**, γ-TeO$_2$ (space group $P2_12_12_1$) and **d**, tellurite β-TeO$_2$ (space group $Pbca$).

The synthetic, colorless tetragonal α-TeO$_2$ (paratellurite) under ambient conditions was previously reported to exhibit a high-symmetry rutile-type structure (space group $P4_2/mnm$),[1] where each Te atom is coordinated by six O atoms, with a Te—O bond length of 2.30 Å, as shown in supplementary Fig. 1a. Subsequent studies[1] revised the structure of α-TeO$_2$ to a distorted rutile-type with space group $P4_12_12$ and a doubled $c$ axis, as depicted in supplementary Fig. 1b. Despite this distortion, each Te atom remains 6-coordinated, forming a highly distorted octahedron, where the two short "quasi-equatorial" bonds, two short non-equatorial bonds, and two long non-equatorial bonds are 2.12 Å, 1.88 Å, and 2.87 Å, respectively. More often than not, the two long bonds are excluded, and Te is considered 4-coordinated. Thus, α-TeO$_2$ can be described as a three-dimensional network of corner-sharing TeO$_4$ or TeO$_4E$ ($E$ represents the lone pair of Te) disphenoids. At high pressures above 0.91 GPa,[2] α-TeO$_2$ transforms into a lower-symmetry phase (referred to as γ-TeO$_2$), which belongs to the orthorhombic space group $P2_12_12_1$, as illustrated in supplementary Fig. 1c. This phase transition involves the splitting of the tetragonal $a$ lattice parameter into the unequal $a$ and $b$ orthorhombic parameters, causing the splitting of O sites (denoted as O1 and O2) and resulting in varied bond lengths. The four shorter Te—O bond lengths range from 1.86 to 2.20 Å, while the two longer Te—O bond lengths are 2.69 Å and 3.20 Å, respectively. On the other hand, as illustrated in supplementary Fig. 1d, the naturally occurring yellow mineral tellurite (β-TeO$_2$) crystallizes in an orthorhombic layered structure,[3,4] which crystallographically resembles brookite, the orthorhombic variant of TiO$_2$. Unlike the three-dimensional networks of α-TeO$_2$ and γ-TeO$_2$, in which the TeO$_4$ disphenoids are corner-sharing, the two-dimensional layers in β-TeO$_2$ consist of TeO$_4$ disphenoids by sharing alternately the O corners and the O—O edges. Each layer contains two types of O positions in: those inside the layer (labeled O1) and those on the layer's surface (labeled O2). When β-TeO$_2$ consists of only one or a few layers, the so-called 2D β-TeO$_2$ is obtained.[5,6]

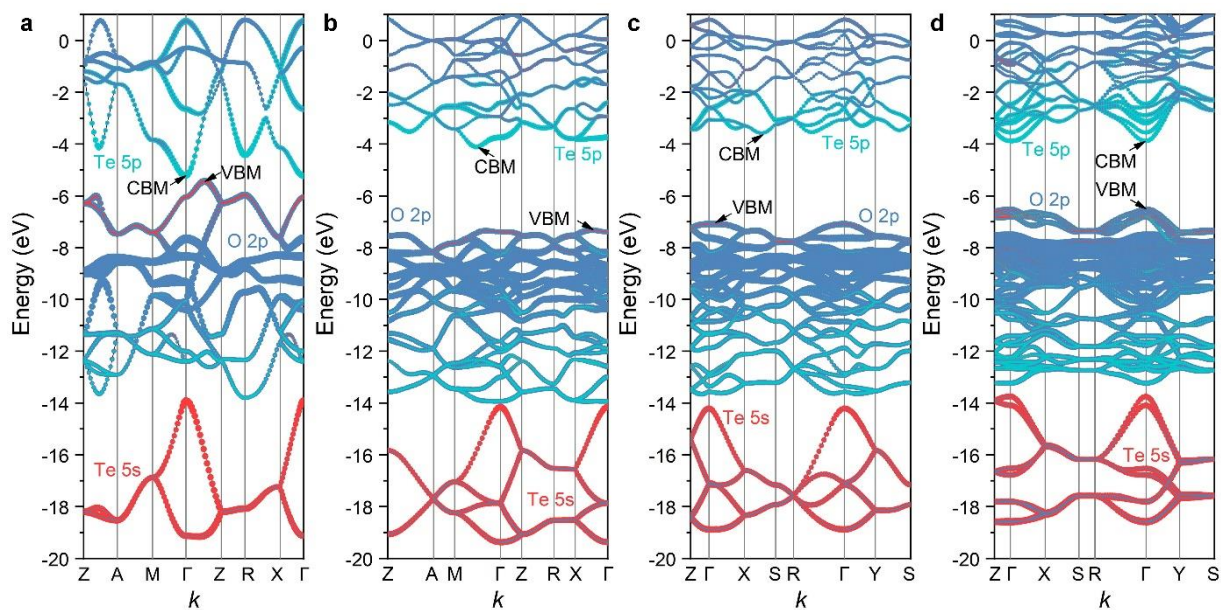

**Supplementary Fig. 2 | Calculated band structures of TeO$_2$ polymorphs: a**, hypothetical rutile-type TeO$_2$ (space group $P4_2/mnm$), **b**, paratellurite α-TeO$_2$ (space group $P4_12_12$), **c**, γ-TeO$_2$ (space group $P2_12_12_1$) and **d**, tellurite β-TeO$_2$ (space group $Pbca$). The energies are with respect the vacuum level (zero energy).

For the hypothetical rutile-type TiO$_2$, the calculated band structure is shown in Supplementary Fig. 2a. The structure reveals an indirect bandgap of only 0.20 eV, which may explain the metallic behavior of recently-reported r-TiO$_2$ thin films on the FeTe surface.[7] The VBM consists of antibonding states of O 2p and Te 5s orbitals, while CBM is predominantly derived from Te 5p orbitals. Despite the substantial energy difference between Te 5s and O 2p orbitals (−15.1 eV vs. −9.0 eV, see Supplementary Table 3), the high symmetry of r-TiO$_2$ results in a significant coupling between these orbitals, resulting in a notably elevated VBM. Moreover, the high symmetry permits excellent dispersion of the conduction band derived from Te 5p orbitals. As r-TeO$_2$ distorts into α-TeO$_2$, shown in Supplementary Fig. 2b, the coupling between Te 5s and O 2p orbitals weakens dramatically, leading the VBM to be primarily composed of O 2p orbitals with minor contribution from the Te 5s states. The structural distortion also narrows the conduction band derived from Te 5p orbitals, widening the calculated bandgap of α-TeO$_2$ to 3.21 eV, which align with optical transparency. For the high-pressure orthorhombic γ-TeO$_2$ (Supplementary Fig. 2c), its band structure is similar to that of tetragonal α-TeO$_2$, but with a slightly larger bandgap, likely attributed to further symmetry reduction and bond breaking from α- to γ-TeO$_2$. In contrast, β-TeO$_2$ (Supplementary Fig. 2d) exhibits a markedly different band structure: (i) It has a direct bandgap at the Γ point, unlike the indirect bandgaps of α-TeO$_2$ and γ-TeO$_2$; (ii) The bandgap of β-TeO$_2$ is 2.63 eV, significant smaller than those of α- and γ-TeO$_2$, correlating with its coloration; (iii) The band edges of β-TeO$_2$ exhibit high anisotropy, with pronounced dispersion along the Γ–Y and Γ–X directions and localized behavior along the Γ–Z direction, reflecting its layered structure.

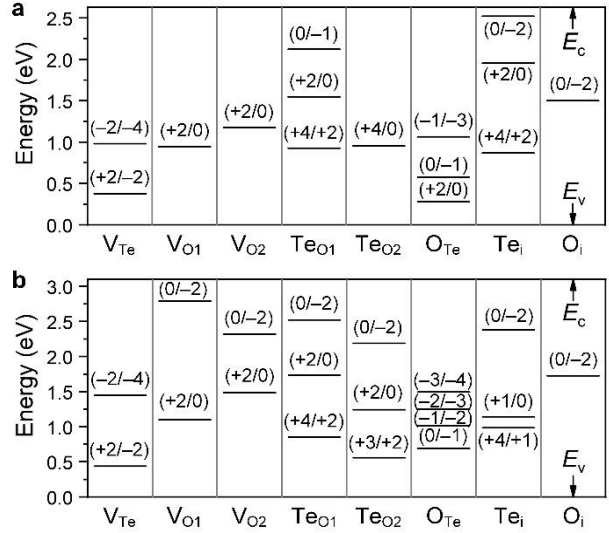

**Supplementary Fig. 3 | Calculated defect transition levels: a**, 3D β-TeO$_2$ bulk, **b**, 2D β-TeO$_2$ monolayer. "$E_v$" and "$E_c$" represent the valence band maximum and conduction band minimum, respectively.

Similar to PbO,[8] all intrinsic defects in β-TeO$_2$ exhibit deep transition levels (Supplementary Fig. 3a). Specifically, the two distinct oxygen vacancies (V$_{O1}$ and V$_{O2}$) show (+2/0) transitions at 1.69 eV and 1.42 eV below the CBM, respectively, indicating they exist in neutral states when $E_F$ is close to CBM. On the other hand, the calculated (0/–2) transition of Te vacancy (V$_{Te}$) is at 0.36 eV above the VBM. This deep transition is explained by the insufficient contribution of Te 5s$^2$ orbital to the VBM. Notably, the V$_{Te}$ is a negative U system, exhibits a (+2/–2) transition at 0.38 eV above the VBM, so the V$_{Te}$ acts not as an acceptor but rather as a hole trap when the $E_F$ is in the $p$-type region. The Te interstitial (Te$_i$) possesses a (0/+2) transition at 0.67 eV below the CBM, while the oxygen interstitial (O$_i$) shows a (0/–2) transition at 1.50 eV above the VBM, indicating that they are ineffective as donors and acceptors, respectively. The three antisite defects (denoted as Te$_{O1}$, Te$_{O2}$, and O$_{Te}$) can be considered as complexes involving the corresponding vacancies and interstitials, and as such, they also exhibit deep transition levels. Compared to 3D β-TeO$_2$, the intrinsic defects in 2D β-TeO$_2$ essentially shows similar deep transition levels (Supplementary Fig. 3b).

**Supplementary Table 1 | Data for band alignment.** Experimental electron affinities ($\chi$), ionization potentials ($IP$), and bandgaps ($E_g$) of typical $n$-type TCOs (such as ZnO, In$_2$O$_3$, and SnO$_2$), typical $p$-type TCOs (such as SnO, NiO, and CuAlO$_2$) and compounds considered but found challenging to dope as $p$-type (such as PbO and CsPbCl$_3$)

| Compound | Measured parameters | Derived parameters |
| --- | --- | --- |
| ZnO | $\chi$: 4.35 eV;[9] $E_g$: 3.3 eV[9] | $IP = \chi + E_g$ = 7.65 eV |
| In$_2$O$_3$[a] | $IP$: 7.10 eV;[10] $E_g$: 2.63 eV[11] (3.6 eV[11]) | $\chi = IP - E_g$ = 4.47 eV (3.5 eV) |
| SnO$_2$ | $\chi$: 4.53 eV;[12] 3.6 eV[13] | $IP = \chi + E_g$ = 8.13 eV |
| SnO[a] | $IP$: 5.8 eV;[14] $E_g$: 0.7 eV[14] (2.7 eV[14]) | $\chi = IP - E_g$ = 5.1 eV (3.1 eV) |
| NiO | $IP$: 5.4 eV;[15] $E_g$: 3.7 eV[15] | $\chi = IP - E_g$ = 1.7 eV |
| CuAlO$_2$[a] | $IP$: 5.24 eV;[16] $E_g$: 2.99 eV[17] (5.33 eV[17]) | $\chi = IP - E_g$ = 2.25 eV (1.71 eV) |
| PbO | $IP$: 6.2 eV;[8] $E_g$: 2.8 eV[8] | $\chi = IP - E_g$ = 3.4 eV |

[a]For indirect bandgap semiconductors, the direct bandgap values are all provided in parentheses; correspondingly, the "$\chi$" values derived using the direct bandgaps are also given in parentheses following $\chi$.

**Supplementary Table 2 | Atomic orbital energies (eV) relative to the vacuum level.**

| Element | O | Sn | Pb | Sb | Bi | Te | Po |
|---|---|---|---|---|---|---|---|
| Outermost p orbital | 2p: −9.0 | 5p: −3.7 | 6p: −3.5 | 5p: −4.8 | 6p: −4.5 | 5p: −5.9 | 6p: −5.6 |
| Outermost s orbital | 2s: −23.9 | 5s: −10.5 | 6s: −12.0 | 5s: −12.8 | 6s: −14.4 | 5s: −15.1 | 6s: −16.9 |

**Supplementary Table 3 | Calculated electrical parameters.** Calculated equilibrium Fermi level ($E_{F,e}$) relative to the VBM, electron density ($n_0$), and hole density ($p_0$) as a function of the growth temperature ($T_G$) for 3D β-TeO$_2$ bulk and 2D β-TeO$_2$ monolayer under Te-rich and O-rich conditions. The unit of $T_G$ is K, the unit of $E_{F,e}$ is eV, and the units of $n_0$ and $p_0$ are cm$^{-3}$ for 3D β-TeO$_2$ and cm$^{-2}$ for 2D β-TeO$_2$.

| $T_G$ | 3D β-TeO$_2$ | | | | | | 2D β-TeO$_2$ | | | | | |
|---|---|---|---|---|---|---|---|---|---|---|---|---|
| | Te-rich | | | O-rich | | | Te-rich | | | O-rich | | |
| | $E_{F,e}$ | $n_0$ | $p_0$ | $E_{F,e}$ | $n_0$ | $p_0$ | $E_{F,e}$ | $n_0$ | $p_0$ | $E_{F,e}$ | $n_0$ | $p_0$ |
| 300 | 1.37 | 9.5E-05 | 5.5E-03 | 1.25 | 8.1E-03 | 6.4E-05 | 1.71 | 1.1E-16 | 1.5E-11 | 1.38 | 3.1E-11 | 5.3E-17 |
| 350 | 1.42 | 1.1E-05 | 4.7E-02 | 1.06 | 1.3E+01 | 4.1E-08 | 1.76 | 1.6E-17 | 1.0E-10 | 1.34 | 1.5E-10 | 1.1E-17 |
| 400 | 1.47 | 2.1E-06 | 2.4E-01 | 0.94 | 1.7E+03 | 3.1E-10 | 1.74 | 3.5E-17 | 4.7E-11 | 1.36 | 6.7E-11 | 2.5E-17 |
| 450 | 1.50 | 6.0E-07 | 8.6E-01 | 0.83 | 1.2E+05 | 4.2E-12 | 1.72 | 7.7E-17 | 2.1E-11 | 1.41 | 1.2E-11 | 1.4E-16 |
| 500 | 1.53 | 1.8E-07 | 2.8E+00 | 0.74 | 3.9E+06 | 1.3E-13 | 1.70 | 1.3E-16 | 1.3E-11 | 1.44 | 3.0E-12 | 5.5E-16 |
| 550 | 1.62 | 6.0E-09 | 8.6E+01 | 0.66 | 6.8E+07 | 7.6E-15 | 1.69 | 2.0E-16 | 8.3E-12 | 1.47 | 1.1E-12 | 1.5E-15 |
| 600 | 1.71 | 1.6E-10 | 3.1E+03 | 0.60 | 7.3E+08 | 7.0E-16 | 1.68 | 2.8E-16 | 6.0E-12 | 1.49 | 5.5E-13 | 3.0E-15 |
| 650 | 1.79 | 8.0E-12 | 6.4E+04 | 0.55 | 5.5E+09 | 9.4E-17 | 1.68 | 3.7E-16 | 4.5E-12 | 1.50 | 3.4E-13 | 4.8E-15 |
| 700 | 1.85 | 6.3E-13 | 8.2E+05 | 0.50 | 3.1E+10 | 1.7E-17 | 1.67 | 4.6E-16 | 3.6E-12 | 1.51 | 2.4E-13 | 6.9E-15 |

**Computational methods**

First-principles calculations were conducted in the framework of density functional theory using the projector-augmented wave method implemented within the Vienna Ab initio Simulation Package (VASP 6.4.3 code).[18] A plane-wave cutoff energy of 500 eV was employed. For TeO$_2$ polymorphs and their competing phases (Te, Te$_4$O$_9$, Te$_2$O$_5$, TeO$_3$, and O$_2$), Γ-centered $k$-meshes with a $k$-point spacing of 0.2 Å$^{-1}$ were utilized to sample the Brillouin zones. The crystal or molecular structures were fully relaxed using the PBE functional[19] until the forces on each atom were less than 0.01 eV/Å. The PBE[19] functional, known for significantly underestimating the bandgaps for oxides, slightly underestimates those of TeO$_2$ polymorphs. Specifically, the PBE-calculated bandgaps for α- and γ-TeO$_2$ are 2.79 eV and 3.06 eV, respectively, which are relatively close to their experimental values (3.48 eV[20] and 3.41 eV,[21] respectively). Conversely, the HSE[22,23] hybrid functional with a standard mixing parameter of 0.25 (HSE$^{0.25}$), overestimates these bandgaps. The HSE$^{0.25}$-calculated bandgaps for α- and γ-TeO$_2$ are 3.89 eV and 4.21 eV, notably higher than experimental values. Similarly, for β-TeO$_2$, which is yellow, the HSE$^{0.25}$ predicts a bandgap approximately 3.3 eV, inconsistent with its observed color. To gain more accurate bandgaps for TeO$_2$ polymorphs, the HSE functional with a reduced mixing parameter of 0.10 (HSE$^{0.10}$) was examined. The HSE$^{0.10}$-calculated bandgaps for α- and γ-TeO$_2$ are 3.21 eV and 3.51 eV, closely aligning with experimental values, while the HSE$^{0.10}$ bandgap for β-TeO$_2$ is 2.63 eV, consistent with its yellow appearance.[3,4] Consequently, the mixing parameter was optimized to 0.10 to roughly reproduce the experimental bandgaps and this value is applied for electronic structure and total energy calculations.

In the case of the 2D β-TeO$_2$ monolayer and multilayers, the slab models were fully relaxed with the $c$ lattice parameter constrained. For the β-TeO$_2$ monolayer, the $c$ lattice parameter of the slab model was constrained to 12.255 Å, corresponding to the relaxed $c$ lattice parameter of bulk β-TeO$_2$. The thickness of the included vacuum layer (approximately 8.2 Å) has been confirmed to provide reasonably convergent results (cf. calculated bandgaps of 2.667 eV, 2.660 eV and 2.674 eV for vacuum thicknesses of 8.2 Å, 13.9 Å and 20.9 Å, respectively).

The CBMs and VBMs (with respect to the vacuum level) of the TeO$_2$ polymorphs were determined by the methodology described in Ref. 24. (001) surface models with vacuum thickness of 15 Å were used. For bulk α-, β-, and γ-TeO$_2$ polymorphs, the slab thickness were 15–27 Å. The ionization potential (IP) and electron affinity ($\chi$) were calculated using a bulk-based definition via electrostatic alignment between the surface and the bulk as

$$\varepsilon_{IP} = \Delta\varepsilon_{vac-ref} - \Delta\varepsilon_{VBM-ref}$$
$$\varepsilon_{\chi} = \Delta\varepsilon_{vac-ref} - \Delta\varepsilon_{CBM-ref}$$
(1)

where $\Delta\varepsilon_{vac-ref}$ is the energy difference between the electrostatic potential in the vacuum region, namely the vacuum level, and the reference level in a bulk-like region of a surface supercell. $\Delta\varepsilon_{VBM-ref}$ and $\Delta\varepsilon_{CBM-ref}$ are the energy differences between the VBM and the reference level and between CBM and the reference level, respectively. The VBMs and CBMs are aligned using the negatives of the IP and the $\chi$, respectively.

For β-TeO$_2$ bulk and monolayer, defect calculations were carried out using 2×2×1 supercells with an effective size (the cube root of the supercell volume) of approximately 11.6 Å (containing 96 and 48 atoms, respectively) and a single Γ k-point. Using the oxygen vacancies in the β-TeO$_2$ monolayer as an example, a convergence test indicates that the errors in defect formation energies between the employed supercell and a larger supercell (a 2×2×1 supercell with $c$ = 18 Å and an effective size of 17.6 Å) are within 0.1 eV (e.g., for V$_{O2}^{2+}$ in the β-TeO$_2$ monolayer, the $\Delta H$ values at $E_F = E_v$ are −1.85 eV and −1.80 eV for supercell sizes of 11.6 Å and 17.6 Å, respectively), which is considered sufficient convergence for the purpose of this work. The lattice parameters of the defective supercell were constrained, while the atomic positions were relaxed using the PBE functional until the forces on each atom were less than 0.03 eV/Å. The total energies were calculated using the HSE$^{0.10}$.

For a defect ($\alpha$) in a charge state $q$, the formation energy $\Delta H_{\alpha,q}$ was calculated by the equation[25]

$$\Delta H_{\alpha,q} = E_{\alpha,q} - E_h + q(E_v + E_F) + \sum n_i \mu_i, \quad (2)$$

where $E_{\alpha,q}$ is the total energy of the supercell with the defect ($\alpha$) in the charge $q$ obtained by the self-consistent potential correction (SCPC) method[26] implemented in the VASP 6.4.3 package and $E_h$ is that of the perfect host supercell. $E_F$ is the Fermi level referred to VBM, $E_V$. $n_i$ indicates the number of $i$ atom added ($n_i < 0$) or removed ($n_i > 0$) when a defect is formed, and $\mu_i$ is the chemical potential of the $i$ atom which can be expressed with respect to that of an element phase ($\mu_i^{el}$) by $\mu_i = \mu_i^{el} + \Delta\mu_i$.

The charge transition level $\varepsilon(q/q')$ was calculated by the equation

$$\varepsilon(q/q') = \frac{\Delta H_{\alpha,q'} - \Delta H_{\alpha,q}}{q - q'}, \quad (3)$$

where $\Delta H_{\alpha,q}$ and $\Delta H_{\alpha,q'}$ are the formation energies of a defect $\alpha$ at the charge states $q$ and $q'$, respectively.

The defect density was calculated by the statistic equation

$$c_{\alpha,q}(E_F, \mu, T_G) = N_{\alpha,q} \exp\left[\frac{-c_{\alpha,q}(E_F, \mu)}{k_B T_G}\right], \quad (4)$$

where $N_{\alpha,q}$ is density of possible sites for defects, $k_B$ is the Boltzmann constant, and $T_G$ is the growth temperature at which defects are formed and assumed to be frozen at room temperature (300 K). The equilibrium $E_F$ ($E_{F,e}$) was determined by solving the following semiconductor statistic equations self-consistently to satisfy the charge neutrality condition:[27]

$$\sum_j \sum_k q_k c_{\alpha_j,q_k} - n_0 + p_0 = 0, \quad (5)$$

$$\sum_j \sum_k q_k c_{\alpha_j,q_k} - n_0 + p_0 = 0$$

$$n_0 = 2\left(\frac{2\pi m_e^* k_B T}{h^2}\right)^{3/2} \exp\left[\frac{-(E_c - E_F)}{k_B T}\right], \quad (6)$$

$$p_0 = 2\left(\frac{2\pi m_h^* k_B T}{h^2}\right)^{3/2} \exp\left[\frac{-(E_F - E_v)}{k_B T}\right], \tag{7}$$

where $E_c$ and $E_v$ are the CBM and VBM, respectively; $m_e^*$ and $m_h^*$ are effective masses of electrons and holes, respectively; $h$ is the Planck constant; $T$ is the temperature for measuring electrical properties.